\documentclass[conference]{IEEEtran}
\IEEEoverridecommandlockouts
\usepackage{cite}
\usepackage{url}
\usepackage{amsmath,amssymb,amsfonts}
\usepackage{algorithmicx}
\usepackage{graphicx}
\usepackage{textcomp}
\usepackage{balance}
\usepackage{xcolor}
\def\BibTeX{{\rm B\kern-.05em{\sc i\kern-.025em b}\kern-.08em
    T\kern-.1667em\lower.7ex\hbox{E}\kern-.125emX}}

\usepackage{algorithm}
\usepackage[noend]{algpseudocode}

\usepackage{listings}
\usepackage{multirow} 

\begin{document}

\title{PULSE: Parametric Hardware Units for Low-power Sparsity-Aware Convolution Engine
}

\author{Ilkin Aliyev and Tosiron Adegbija\\
Department of Electrical and Computer Engineering\\
The University of Arizona, Tucson, AZ, USA\\
Email: \{ilkina, tosiron\}@arizona.edu

}

\maketitle

\begin{abstract}

Spiking Neural Networks (SNNs) have become popular for their more bio-realistic behavior than Artificial Neural Networks (ANNs). However, effectively leveraging the intrinsic, unstructured sparsity of SNNs in hardware is challenging, especially due to the variability in sparsity across network layers. This variability depends on several factors, including the input dataset, encoding scheme, and neuron model. Most existing SNN accelerators fail to account for the layer-specific workloads of an application (model + dataset), leading to high energy consumption. To address this, we propose a design-time parametric hardware generator that takes layer-wise sparsity and the number of processing elements as inputs and synthesizes the corresponding hardware. The proposed design compresses sparse spike trains using a priority encoder and efficiently shifts the activations across the network’s layers. We demonstrate the robustness of our proposed approach by first profiling a given application's characteristics followed by performing efficient resource allocation. Results on a Xilinx Kintex FPGA (Field Programmable Gate Arrays) using MNIST, FashionMNIST, and SVHN datasets show a $3.14\times$ improvement in accelerator efficiency (FPS/W) compared to a sparsity-oblivious systolic array-based accelerator. Compared to the most recent sparsity-aware work, our solution improves efficiency by $1.72\times$. 

\end{abstract}

\begin{IEEEkeywords}
Sparsity-aware SNN, FPGA, parametric hardware design, neuromorphic computing.
\end{IEEEkeywords}

\section{Introduction}
Spiking Neural Networks (SNN) are emerging as a competitive alternative to traditional Artificial Neural Networks (ANNs) due to the biologically plausible neuron model (i.e. binary activation and multiplication-free) which operates on sparse activations. However, most studies in SNN hardware literature ignore this characteristic despite its significant impact on hardware performance. As evidence of the potential for optimization, prior work \cite{yin2022sata} improved training energy by $5.58\times$ compared to SNNs that did not exploit sparsity. Similarly, Wang et. al. \cite{wang2023spiking} improved inference delay by $2.1\times$ compared to sparsity-oblivious hardware.

To fully realize the efficiency benefits of SNNs, the hardware implementations must match the SNN's computation needs. A key aspect of this involves determining the number of processing elements (PEs) for deep learning (DL) accelerators, as this significantly influences resource utilization and energy consumption. This ongoing research challenge has prompted various approaches \cite{venkatesan2019magnet}, such as SCNN's configuration of 1024 Multiply-and-Accumulate (MAC) units for HD video processing \cite{parashar2017scnn} and Eyerissv2's exploration of diverse models with different layer shapes \cite{eyeriss2019_jetcas}. Additionally, it is crucial to effectively and explicitly exploit the sparsity characteristic of SNNs. This requires rethinking traditional methods used in conventional convolutional neural networks (CNNs), like the shifting window-based approach \cite{shen2023nbssn, liu2023low}, due to the variable sparsity across network layers, feature maps (fmaps), and within each fmap's timesteps. Addressing this complexity is essential for meeting the latency, area, and memory requirements of complex SNNs like spiking CNNs.    

\begin{figure}[t!]
\vspace{-5pt}
		\centering
		\includegraphics[width=0.85\linewidth]{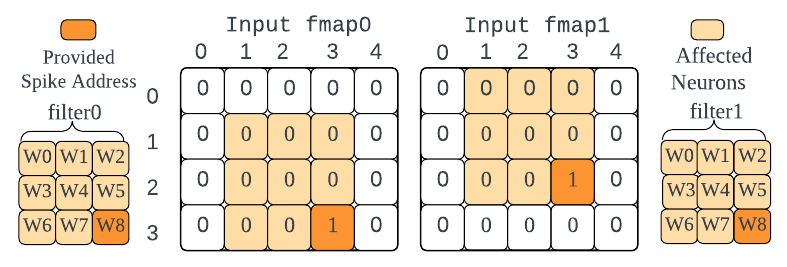}
		\vspace{-10pt}
		\caption{Illustration of spike-based convolution operation flow. The ``membrane potential" of neurons with non-zero activation values is updated along with the surrounding neurons determined by the filter weights.}
		\label{fig:sparse_conv_illustr}
		\vspace{-15pt}
\end{figure}

\noindent\textbf{Event-driven SNN hardware}: Figure \ref{fig:sparse_conv_illustr} illustrates the concept of convolution (CONV) operations on sparse activations. Unlike classical CONV operations where the filter is slid over the input activations, the spike event addresses in the input activations determine which neurons in the post-synaptic layer will be updated. 
We argue that by specializing hardware resource allocation to the sparsity of each layer, we can reduce excessive resource demands and optimize the efficiency of SNN accelerators. Recent studies \cite{leigh2022selective, leigh2023digital, aliyev2023design} have shown reductions in both hardware resources and inference time by simply considering sparsity in the input layer in a two-layer network with only input and output layers.

Prior sparsity-aware studies have two main limitations. Firstly, works using systolic array-based hardware follow a ``one size fits all" approach to sparsity exploitation across layers \cite{sommer2022efficient}. Consequently, while small layers under-utilize resources, leaving PEs idle and causing increased leakage power, larger layers overutilize resources causing latency overhead. Drastically unbalanced sparsity across the layers necessitates a layer-wise resource specialization. Secondly, network-on-chip (NoC)-based event-driven designs \cite{davies2018loihi, di2022sne} face substantial runtime overhead due to managing (routing and storing) large spike trains, impacting the system's energy efficiency. As a result, the non-trivial problem of mapping the model onto hardware \cite{eyeriss2019_jetcas} becomes even more cumbersome due to sparsity. 

To alleviate these issues, we propose \textit{\textbf{PULSE} (Parametric Hardware Units for Low-power
Sparsity-Aware Convolution Engine)}, a novel event-driven modular hardware design that accelerates sparsely active convolution operations. PULSE is a fully parametrized hardware design that can address the flexibility and efficiency needs of different spiking CNN models. We employ design time parameters to partition hardware resources across network layers such that the \textit{layer-wise} workload (i.e., input activation sparsity, layer size) is taken into account. 
Implementation results on a Xilinx Kintex UltraScale FPGA (Field Programmable Gate Arrays) using three datasets---MNIST, Fashion MNIST, and Street View House Numbers (SVHN)---show that PULSE can achieve $1.73\times$ and $3.14\times$ better frames per second per watt (FPS/W) accelerator efficiency compared to sparsity-aware \cite{sommer2022efficient} and sparsity-oblivious \cite{ye2022implementation} studies, respectively.  

\begin{algorithm}[!t]
\caption{Even-based Convolution Loop Flow}
\begin{algorithmic}[1]
    \For{each output channel}
    \Comment{$\triangleright$ \textcolor{red}{unrollable}}
        \For{each time step}
            \For{each input channel}
                \For{each spike event}
                    \For{each kernel column} \Comment{$\triangleright$ \textcolor{red}{unrollable}}
                        \For{each kernel row} \Comment{$\triangleright$ \textcolor{red}{unrollable}}
                        \State Calculate affected neuron index
                        \State Update neuron membrane potential 
                        \hspace*{9em} with the corresponding filter weight
                    \EndFor
                \EndFor 
            \EndFor
        \EndFor
        \For{each neuron's membrane pot. in OFM}
            \State Add filter bias to it
            \State Perform comparison \& thresholding on it
            \State Apply leakage to it
        \EndFor
    \EndFor
\EndFor
\end{algorithmic}
\end{algorithm}

\section{Related Work}
To contextualize our work with the state-of-the-art, we briefly highlight two prior studies that provide sparse convolution engines. Sommer et al. \cite{sommer2022efficient} propose an FPGA-based SNN accelerator with at least 9 PEs, each PE handling one filter coefficient add operation, and utilized "Memory Interlacing" for parallel access to neurons' membrane potentials. Their design also supports parallelization of output channels, allowing deployment of multiple 9 PE clusters to handle various output feature maps in a given CONV layer. Although we employ output channel-wise parallelization, our approach differs by not parallelizing filter coefficients to simplify memory management (details in Section \ref{sec:workload_balance}). 
Mauro et al. \cite{di2022sne} propose an ASIC implementation of a NoC-based hardware capable of handling entire CNN layers on a single chip. Each PE in their NoC handles 16 hardware neuron instances, with custom packet routing logic for output spikes. While their results show the NoC's flexibility in exploiting SNN layers' irregular dataflow, it incurs significant routing delays of the encoded spike packets and requires additional hardware logic in each PE for packet handling, increasing the hardware complexity.

\section{Balancing sparse workload} \label{sec:workload_balance}
The primary challenge in implementing a parallelization strategy in sparsity-aware SNN accelerators is maintaining the continuous operation of hardware units to prevent the associated power overheads of idle units. Algorithm 1 shows the convolution loop flow carried out in an event-based fashion. The algorithm also identifies the loops that can be unrolled without causing data hazards, like conflicts that arise when multiple feature maps write to the same output channel. 

To identify the best parallelization strategy for event-based spiking convolution, we examine various options in Algorithm 1. Firstly, input channel-wise parallelization (lines 3 and 4) is inefficient due to fluctuating activation sparsity in each input feature map (IFM), leading to workload imbalance among the PEs, and data hazards when IFMs contribute to the same output feature map (OFM) with overlapping spike events. Secondly, while kernel weight parallelization (lines 5 and 6) avoids these issues as it involves updating independent neurons' membrane potentials (see Figure \ref{fig:sparse_conv_illustr}), it requires on-chip parallel memory access. Sommer et. al. \cite{sommer2022efficient} address this challenge through memory interlacing, but at the cost of hardware overheads. Lastly, output channel-wise parallelization (line 1) appears most effective, as each PE can independently process the same spike events and update its neurons. 

However, output channel-wise parallelization comes at the expense of additional on-chip memory since each PE needs an OFM-sized memory to keep track of neuron membrane potentials. Furthermore, additional comparator and subtraction logic are needed for the activation phase of the leaky
integrate-and-fire (LIF) neuron. The latter can be efficiently circumvented by employing low-cost OR gates instead of a full 32-bit fixed point comparison (see Section \ref{sec:pulse_arch}). We argue that the former can be addressed by further chunking the OFM and assigning a fraction of activations to each parallel PE. Once each PE completes all time steps and all spikes, it moves on to the next spatial chunk of the OFM. As a result, we can reduce the on-chip memory required for membrane potential storage, while maintaining the parallelization degree and, most notably, avoiding the hardware cost of managing interlacing.

The characteristics of the LIF neuron, used in this work, are shown in Equations \ref{ref:lif1} and \ref{ref:lif2}. 

\vspace{-3mm}
\begin{equation} \label{ref:lif1}
    u_{j}[t+1] = \beta u_{j}[t] + \sum_{i} w_{ij} s_{i}[t] - s_{j}[t] \theta
\end{equation}
\vspace{-3mm}
\begin{equation}\label{ref:lif2}
s_{j}[t] = 
\begin{cases} 
1, & \text{if } u_{j}[t] > \theta \\
0, & \text{otherwise}
\end{cases}
\end{equation}

\noindent where, $u_{j}[t+1]$ represents the membrane potential for previous and current time steps, $\beta$ is the decay, and $\theta$ is the threshold value that the membrane potential must surpass to trigger a firing event, as indicated by a spike $s_{j}[t]$.

\section{PULSE Architecture} \label{sec:pulse_arch}
\begin{figure}[t!]
\vspace{-5pt}
		\centering
		\includegraphics[width=0.8\linewidth]{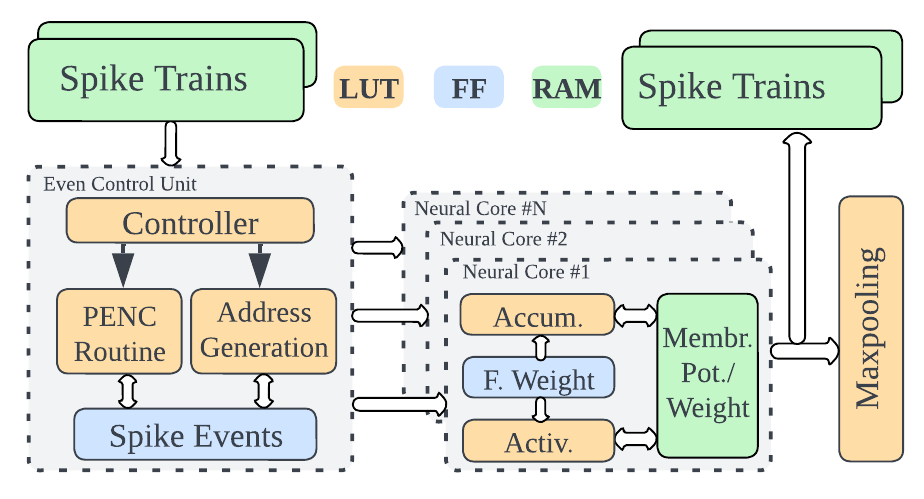}
		\vspace{-15pt}
		\caption{The proposed layer architecture for a CONV/FC layer hardware. PENC stands for Priority Encoder routine. F. weight stands for Filter weights}
		\label{fig:arch}
		\vspace{-15pt}
\end{figure}

Figure \ref{fig:arch} depicts PULSE's hardware architecture, which has two main components: \textit{Event Control Unit (ECU)} and \textit{Neural Core (NC)}. This design unrolls the output channels by a factor of $N$, defined at design time, to determine the number of \textit{NC} instances. Within the \textit{ECU}, the \textit{controller} submodule implements lines 1-6 in Algorithm 1, striding through output channels by $N$. The \textit{controller}'s signals are connected to all parallel \textit{NCs} and interpreted by the \textit{NCs} as a base address. This base address is then paired with an offset parameter with which each \textit{NC} is instantiated to calculate the OFM index. For instance, an \textit{NC} with an offset of 2 and $N$=8, will process OFMs with indices 1, 9, 17, etc. 

The \textit{controller} fetches a spike train from the input Spike RAM, then initiates the Priority Encoder (PENC) routine. This routine compresses the spike train by eliminating the non-spiking (i.e., `0') bits, translating an n-bit spike train into a register array, $Spike Events$ (see Figure \ref{fig:arch}). PENC processes $n$ bits in each cycle, and outputs the address of the first set bit to the $Spike Events$ array. The ECU's bit reset component then resets the bit value of 1 to 0 for this address in the previous cycle's version of the spike train, enabling the PENC to identify the next set bit in the spike train. After compression, the controller begins the accumulation phase, reading from the $Spike Events$ register set and iterating through the filter coefficients (see Figure \ref{fig:sparse_conv_illustr}). For each coefficient, the \textit{Address Generation} routine calculates the $(row,col)$ addresses of neurons associated with that spike event and filter sizes (i.e., for spike event at $(row,col)$, all 9 affected neurons are from $(row-3,col-3)$ to $(row,col)$). These $(row,col)$ signals are connected to the \textit{NCs}. Importantly, the PENC and accumulation phases are overlapped, allowing PENC to process new spike trains simultaneously with the accumulation phase handling the previously compressed spike events.

With the $(row,col)$ pairs provided, the \textit{Accum} routine within each \textit{NC} then performs the accumulation phase of the LIF neuron. It reads the membrane potential value from the Block RAM (BRAM), adds the weight coefficient to it, and writes back the result to the BRAM (line 8, Algorithm 1). Both the \textit{Address Generation} and \textit{Accum} routines are fully pipelined and can update one neuron per cycle. We store the filter weights of the CONV layer in flip flops (FFs). For fully-connected (FC) weight storage, we use Ultra RAMs (URAMs), which have a higher density than BRAMs, enabling a larger storage space and better energy efficiency. Unlike BRAMs, URAMs are fixed in size (e.g., each URAM tile is fixed to 72-bit width and 4K depth). Therefore, we concatenate two FC neuron weights (each weight is 32 bits) in a single row to efficiently utilize the URAM rows. 

The controller, after processing all input feature maps, enables the \textit{Activ.} unit within the \textit{NCs} to start the LIF neuron's spiking phase. This triggers each \textit{NC} to perform lines 9-12 in Algorithm 1 on the output features. Our design uses a simple, low-cost comparator to avoid full 32-bit fixed-point comparisons. We set the threshold value to a constant $1.0$ during training and employ a Q3.29 fixed-point representation in our hardware, with the most significant bit (MSB) as the sign bit. As such, for efficiency, we only check the first 3 bits. If the MSB (bit 31) is 1, then the number is less than $1.0$, while a 1 in the second or third bits indicates a value of $1.0$ or greater, signaling a spiked neuron. Then, thresholding is performed by subtracting $1.0$ from its membrane potential. The architecture also features max pooling, which is effective in SNNs for computer vision datasets. Its implementation on binary feature maps only requires sliding an OR gate over an $N\times N$ input area, where $N$ is the downsampling ratio \cite{sommer2022efficient}.

\section{Implementation Results}

\subsubsection{Experimental Setup} 
We used the Xilinx Kintex\textregistered{} UltraScale+\texttrademark{} FPGA to implement the proposed hardware due to its higher logic resources than Artix and lower power than the Virtex FPGAs. We used \textit{snntorch}\footnote{Available online at \url{https://github.com/jeshraghian/snntorch}} \cite{eshraghian2023training} for training our models using surrogate gradient learning \cite{li2021differentiable}. We used MNIST, Fashion MNIST (FMNIST), and Street View House Numbers (SVHN) datasets as the driving applications. We set the \texttt{beta} hyperparameter to $0.15$ for all datasets and training batch size to $64$. We used two different networks with the following architectures: (1) 32C3-32C3-10C3-MP3-10 and (2) 32C3-P2-32C3-MP2-256-10 (where $X$C$Y$ stands for $X$ filters with size $Y\times Y$ and MP$Z$ for max pooling with size of $Z \times Z$)\footnote{Pre-trained models available at \url{https://github.com/githubofaliyev/model\_zoo}}. We evaluated our accelerator with respect to throughput and efficiency (FPS/W) in comparison to two state-of-the-art (SOTA) SNN accelerators representing sparsity-aware \cite{sommer2022efficient} and regular sparsity-oblivious \cite{ye2022implementation} designs.

We used standard rate coding due to its ability to achieve SOTA accuracy. However, rate codes are known to require lengthy spike trains to achieve high accuracy, and population coding in the classification layer is often effective in reducing the length, at the expense of increased neuron count in the classification layer \cite{temp_vs_spatial_coding}. As such, each class gets mapped to 2 or more neurons (instead of a single neuron) which effectively increases the firing rate of neurons with shorter time steps. This method fits particularly well with our design scheme since the output layer is the most passive in the network. That is, although the neuron count is increased by employing a population of neurons in the classification layer, the highly sparse nature of this layer incurs negligible hardware overhead.

\subsubsection{Results} Table \ref{tab:compr_results} shows the comparison results. It also shows the sparsity range in the dataset images, indicating how sparsity varies even within the same dataset. To determine hardware configurations for each comparison, we calculate the application workload distribution using Equation \ref{res_partit}. Let the workload of a spiking convolutional and fully-connected neural network layer ($W_{\text{CONV}}$ and $W_{\text{FC}}$) be defined as: 
\vspace{-5pt}
\begin{equation} \label{res_partit}
  \begin{aligned}
    W_{\text{CONV}} &= F \times C_{\text{out}} \times \sum_{i=1}^{N} S_i, \quad
    W_{\text{FC}} &= N \times S
  \end{aligned}
\end{equation}

\noindent where \textit{F} is the number of filter coefficients (e.g., $9$ for $3\times3$), \textit{C\_{\text{out}}} is the number of output channels, and $S\_i$ is the number of spikes for input feature map $i$. PULSE is equipped with network-specific performance counters and provides layer-wise execution statistics like the average number of spikes, latency cycles, etc. Therefore, we obtain $S\_i$ by first naively mapping the network onto PULSE with one neural core per layer to reveal each layer's total workload. Then, we refine the hardware to better align with the workload needs.

\begin{figure}[t!]
\vspace{-5pt}
		\centering
		\includegraphics[width=0.9\linewidth]{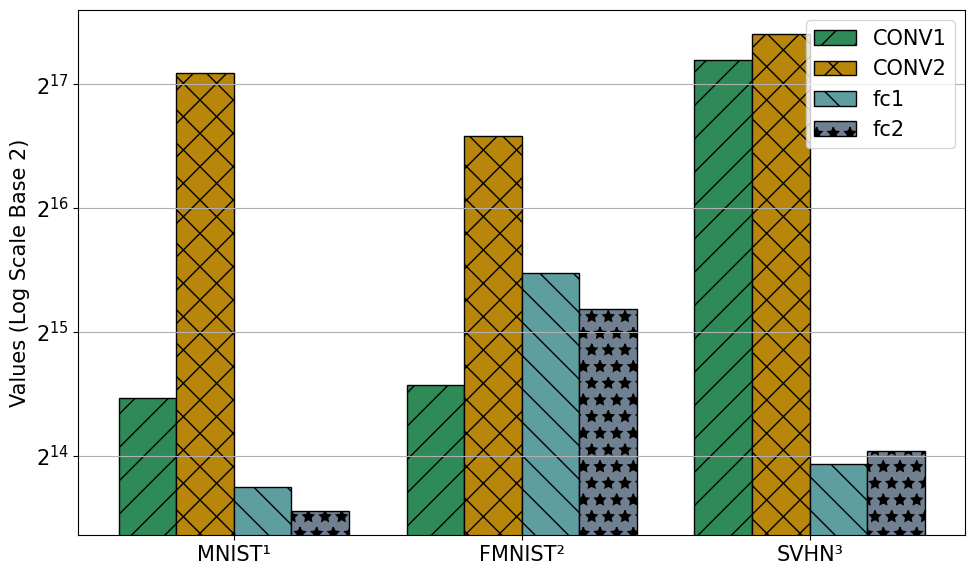}
		\vspace{-10pt}
		\caption{Per-layer workload distribution for each application (model+dataset). The superscripts represent the network used for the dataset, shown below Table \ref{tab:compr_results}.}
		\label{fig:app_wrkl}
		\vspace{-15pt}
\end{figure}

Figure \ref{fig:app_wrkl} shows the layer workload distributions for each application scenario. We observe that the second CONV layer's latency dominates the overall network latency in all three applications due to 32 input features. The most notable gap can be seen in MNIST$^{1}$ because, unlike others, there is no max pooling applied to the CONV2 input. On the other hand, the weight of both CONV1 and CONV2 becomes identical in SVHN$^{3}$ because SVHN images have 3 channels. Based on this analysis, we partitioned the hardware resources to ${8, 32, 4, 2}$ \textit{NCs}, ${16, 32, 8, 8}$, and ${32, 32, 3, 3}$ \textit{NCs} for the four consecutive layers in MNIST$^{1}$,  FMNIST$^{2}$, and SVHN$^{3}$, respectively. This partitioning provides the top throughput with the highest utilization.

\vspace{5pt}
\noindent\textbf{Comparison to previous work:} Compared to the sparsity-aware prior work \cite{sommer2022efficient}, PULSE achieved $1.73\times$ efficiency improvement for MNIST. Prior work outperformed our work in throughput and latency because they utilized TTFS coding, whereas we used the Poisson-based rate coding. However, their power usage is worse due to high frequency, DSP usage, and idle PEs, leading to increased leakage power. Despite a slightly higher area footprint in our work, we achieved a power advantage due to the utilization of URAMs instead of BRAMs. Note that we report power without any power optimizations such as power/clock gating. Compared to \cite{ye2022implementation}, which ignored sparsity, PULSE achieved $3.14\times$ better efficiency for FMNIST$^{2}$. For SVHN$^{3}$, which has lower sparsity ($54\%$ median sparsity), PULSE achieved $2\times$ improvement. 

While PULSE outperformed prior work in efficiency in all cases, it lagged in latency. The latency degraded because each layer needed to finish processing an entire image before moving on to the next layer. Therefore, PULSE favors high-throughput and low-power scenarios. Furthermore, by employing population coding, we significantly reduced the number of time steps without degrading the model accuracy and hence balanced the intensive spiking characteristics of the rate coding for MNIST and FMNIST. For SVHN, we observed accuracy degradations below 18 time steps. Overall, PULSE's innovative hardware design improves efficiency while enabling flexibility for further performance-power tradeoffs.

\begin{table}[t]
    \centering
    \footnotesize
    \vspace{-10pt}
    \caption{Comparison to prior work. The superscripts represent the networks depicted below the table. In the \textit{platform} row, K. stands for Kintex, and Z. stands for Zynq.} \label{tab:compr_results}
    \begin{tabular}{|c||c|c|c|c|c|c|}
        \hline
        \textbf{Dataset} & \multicolumn{2}{c|}{MNIST$^{1}$} & \multicolumn{2}{c|}{FMNIST$^{2}$} & \multicolumn{2}{c|}{SVHN$^{3}$} \\
        \hline \hline
        \textbf{Work} & PULSE & \cite{sommer2022efficient} & PULSE & \cite{ye2022implementation} & PULSE &  \cite{ye2022implementation} \\
        \hline
        \textbf{Latency} & \multirow{2}{*}{131.7K} & \multirow{2}{*}{13.3K} & \multirow{2}{*}{163K} & \multirow{2}{*}{120K} & \multirow{2}{*}{214K} & \multirow{2}{*}{121K} \\
        {[\#cycles]} & & & & & & \\
        \hline
        \textbf{Through-} & \multirow{2}{*}{10.8K} & \multirow{2}{*}{21.4K} & \multirow{2}{*}{2907} & \multirow{2}{*}{833} & \multirow{2}{*}{2155} & \multirow{2}{*}{826} \\
        \textbf{put} {[FPS]} & & & & & & \\
        \hline
        \textbf{LUT}  & 43.3K  & 33K & 47.5K & 80K  & 49.2 & 80K \\
        \textbf{FF}   & 32.6 & 21  & 38.7K & 138K & 40.3 & 138K \\
        \textbf{BRAM} & 42   & 125 & 34   & 246   & 52 & 246 \\
        \textbf{URAM} & 50   & 0   & 53   &   0   & 56  & 0 \\
        \textbf{DSP}  & 0    & 64  & 0    & 0     & 0 & 0 \\
        \hline
        \textbf{Power} & \multirow{2}{*}{0.92} & \multirow{2}{*}{2.9} & \multirow{2}{*}{1.09} & \multirow{2}{*}{0.98} & \multirow{2}{*}{1.28} & \multirow{2}{*}{0.98} \\
        {[W]} & & & & & & \\
        \hline
        \textbf{Efficiency} & \multirow{2}{*}{12.5K} & \multirow{2}{*}{7.2K} & \multirow{2}{*}{2667} & \multirow{2}{*}{848} & \multirow{2}{*}{1684} & \multirow{2}{*}{841} \\
        {[FPS/W]} & & & & & & \\
        \hline
        \textbf{Eff. Gain} & \multicolumn{2}{c|}{$1.73\times$} &  \multicolumn{2}{c|}{$3.14\times$}  & \multicolumn{2}{c|}{$2\times$}   \\
        \hline
        \textbf{Input} & \multicolumn{2}{c|}{high=95} & \multicolumn{2}{c|}{high=76} & \multicolumn{2}{c|}{high=78} \\
        \textbf{Sprs.}{[\%]} & \multicolumn{2}{c|}{low=83} & \multicolumn{2}{c|}{low=68} & \multicolumn{2}{c|}{low=51} \\
        \hline
        \textbf{Input} & \multicolumn{2}{c|}{time step=3} & \multicolumn{2}{c|}{time step=8} & \multicolumn{2}{c|}{time step=18} \\
        \textbf{Coding.} & \multicolumn{2}{c|}{pop. size=500} & \multicolumn{2}{c|}{pop. size=600} & \multicolumn{2}{c|}{pop. size=400} \\
        \hline
        $\textbf{F}_{Max}$ & \multirow{2}{*}{125} & \multirow{2}{*}{333} & \multirow{2}{*}{125} & \multirow{2}{*}{100} & \multirow{2}{*}{125} & \multirow{2}{*}{100} \\
        {[MHz]} & & & & & & \\
        \hline    
        Plat- & \multirow{2}{*}{K. U+} & \multirow{2}{*}{Z. U+} & \multirow{2}{*}{K. U+} & \multirow{2}{*}{K. 7} & \multirow{2}{*}{K. U+} & \multirow{2}{*}{K. 7} \\
        {form} & & & & & & \\
        \hline 
        \textbf{Acc.} {[\%]} & 98.5 & 98.2 & 88.32 & 90.04 & 82.68 & 80.24 \\
        \hline
    \end{tabular}\\
    $^{1}$28x28-32C3-32C3-P3-10C3-10, $^{2}$28x28-32C3-MP2-32C3-MP2-256-10, \\
    $^{3}$32x32x3-32C3-P2-32C3-P2-256-10
    \vspace{-10pt}
\end{table}

\section{Conclusion}
This paper proposes PULSE, an event-driven, \textit{layer-wise}, and flexible hardware design for spiking neural networks. The proposed design mitigates latency overhead by employing \textit{spike-based} processing and an output-channel-wise parallelization strategy. It also reduces power consumption by balancing hardware resources in a layer-wise manner, specializing resources to individual layers' sparsity characteristics. Implementation results on a Kintex FPGA show that PULSE achieves better efficiency (FPS/W) than both sparsity-aware and sparsity-oblivious prior works. Moreover, the paper highlights the challenges inherent in existing SNN hardware designs by analyzing the variability in the sparsity of input datasets, the length of time steps, and the impact of these factors on network latency and hardware efficiency.
Although the factors affecting network latency arise primarily from training, their influence on hardware performance mirrors that of the hardware's design.

\section*{Acknowledgment}
This work was partially supported by the Technology and Research Initiative Fund (TRIF) provided to the University of Arizona by the Arizona Board of Regents (ABOR).

\bibliographystyle{ieeetr}
{\small
\bibliography{refs}}
\balance
\end{document}